\documentclass{article}
\usepackage{booktabs}
\usepackage{tablefootnote}
\usepackage{multirow}
\usepackage{float}
\usepackage{caption}
\usepackage{authblk}
\usepackage[english]{babel}

\usepackage[letterpaper,top=2cm,bottom=2cm,left=3cm,right=3cm,marginparwidth=1.75cm]{geometry}

\usepackage{amsmath}
\usepackage{comment}
\usepackage{graphicx}
\usepackage[colorlinks=true, allcolors=blue]{hyperref}

\title{PC-mer: An Assessment of its Performance in Alignment-Free Phylogenetic Tree Construction}
\author[1]{Saeedeh Akbari Rokn Abadi}
\author[*, 2]{Melika Honarmand}
\author[*, 2]{Ali Hajialinaghi}
\author[1, \string^]{Somayyeh Koohi}
\affil[1]{Sharif University of Technology, Tehran, IRAN}
\affil[2]{Amirkabir University of Technology, Tehran, IRAN}

\date{}
\begin{document}
	\maketitle
	\def\thefootnote{*}\footnotetext{These authors contributed equally to this work}\def\thefootnote{\arabic{footnote}}
	\def\thefootnote{\string^}\footnotetext{Co-author: Koohi@sharif.edu}\def\thefootnote{\arabic{footnote}}

	\begin{abstract}
		
		\noindent\textbf{Background and objective}: Sequence comparison is essential in bioinformatics, serving various purposes such as taxonomy, functional inference, and drug discovery. The traditional method of aligning sequences for comparison is time-consuming, especially with large datasets. To overcome this, alignment-free methods have emerged as an alternative approach, prioritizing comparison scores over alignment itself. These methods directly compare sequences without the need for alignment. However, accurately representing the relationships between sequences is a significant challenge in the design of these tools.

		\noindent\textbf{Methods}:One of the alignment-free comparison approaches utilizes the frequency of fixed-length substrings, known as K-mers, which serves as the foundation for many sequence comparison methods. However, a challenge arises in these methods when increasing the length of the substring (K), as it leads to an exponential growth in the number of possible states. In this work, we explore the PC-mer method, which utilizes a more limited set of words that experience slower growth (${2^{k}}$ instead of ${4^{k}}$) compared to K. We conducted a comparison of sequences and evaluated how the reduced input vector size influenced the performance of the PC-mer method.
		
		\noindent\textbf{Results}: For the evaluation, we selected the Clustal Omega method as our reference approach, alongside three alignment-free methods: kmacs, FFP, and alfpy (word count). These methods also leverage the frequency of K-mers. We applied all five methods to 9 datasets for comprehensive analysis. The results were compared using phylogenetic trees and metrics such as Robinson-Foulds and normalized quartet distance (nQD). 

		\noindent\textbf{Conclusion}: Our findings indicate that, unlike reducing the input features in other alignment-independent methods, the PC-mer method exhibits competitive performance when compared to the aforementioned methods especially when input sequences are very varied.

	\end{abstract}
	\section{Background}
	In the design of sequence comparison tools, two types of output can be produced: similarity/dissimilarity scores and alignments that determine the locations of differences \cite{afprj}. Alignment-based methods calculate and report similarity/dissimilarity scores based on sequence alignment. The scores from this group of methods are often considered accurate and reliable in the field of bioinformatics because they are obtained through the alignment step. However, achieving these scores requires extensive calculations and memory consumption for the alignment process \cite{afprj, vector, fuzzy, walkim}. This is while some applications only require similarity/dissimilarity scores, leading to a desire to use methods that do not require an alignment step to generate scores. Alignment-free methods in sequence comparison have emerged as valuable alternatives to traditional alignment-based approaches. They offer efficient ways to analyze and compare biological sequences without relying on pairwise alignments. By avoiding the computationally intensive alignment step, alignment-free methods can handle large-scale datasets and mitigate the impact of divergent sequences.

Alignment-free methods offer several advantages over alignment-based approaches. Firstly, they are computationally efficient, bypassing the time-consuming alignment step. This efficiency enables the analysis of large datasets within reasonable time frames. Additionally, alignment-free methods are robust against sequence errors or gaps, making them suitable for handling noisy or incomplete data. However, alignment-free methods may have limitations when it comes to capturing fine-grained structural variations or phylogenetic relationships. Alignment-based methods, with their explicit alignment of sequences, excel in detecting local similarities and insertions or deletions (indels). Moreover, alignment-free methods heavily rely on predefined features or sequence representations, omitting potential information that could be derived from full sequence alignments. The choice of an appropriate alignment-free method depends on the specific analysis goals and characteristics of the dataset \cite{afprj}.

Several state-of-the-art alignment-free methods have been developed, each with its unique approach to sequence comparison. One prominent method is the k-mer-based approach, which extracts fixed-length subsequences (k-mers) from input sequences. K-mers are then counted, generating numeric vectors or sequence embeddings that capture essential features of the sequences. Some of notable k-mer-based methods include e.g. CVTree \cite{cvtree}, CAFE \cite{cafe}, and K-mer-SVM \cite{mersvm}. These methods have been used in various applications such as metagenomic classification \cite{met} and studying viruses \cite{vc1, vc2, pcmer} like SARS-CoV-2 \cite{mldspsars, pc2}.

In these methods based on k-mer frequency, there are various methods at different levels of these tools, from counting k-mers to how to analyze the frequency of k-mers. For example, some comparison methods based on k-mer spectrum produce only a certain size of k-mers, such as dinucleotides or trinucleotides \cite{dinuc, tri}, or at level of counting k-mers use the FCGR (Frequence of Chaos Game Representation) method \cite{mldspsars, fcgr}, or even at the level of k-mer spectrum analysis use computing the distance or Discrete Fourier Transformation (DFT) of spectrum \cite{mldspsars, fcgr, dft}. Additionally, supervised machine learning methods, such as support vector machines (SVMs) and random forests, have been integrated with alignment-free approaches to enhance their performance in classification tasks. However, it is essential to consider the significant number of possible states for k-mers (equal to ${4^{k}}$ for DNA and RNA sequences) when using k-mer frequency in machine learning-based structures. For large k values, the input vector's volume grows exponentially, creating challenges in memory consumption and processing \cite{mldspsars, met}.

In conclusion, alignment-free sequence comparison methods, particularly k-mer-based methods, offer valuable alternatives to alignment-based approaches, with their computational efficiency and robustness against errors. However, it is necessary to be mindful of their limitations, including the exponential growth of memory consumption with increasing k values and the accuracy of comparisons based on it. Therefore, this article aims to evaluate the PC-mer method, which is based on k-mers and aims to reduce memory consumption while maintaining or improving the performance of k-mer-based tools \cite{pcmer, pc2}.
	
	The PC-mer function is a representative example of an alignment-free approach for sequence comparison. It utilizes three distinct chemical and physical properties to count k-mers in biological sequences. By counting the occurrences of k-mers based on these properties, the PC-mer function generates informative sequence embeddings.
	
	The PC-mer function's alignment-free nature provides several advantages in the context of sequence comparison and phylogenetic analysis \cite{pcmer, pc2}:
	
	\begin{itemize}
\item[1.] Small and Fixed Feature Space: PC-mer offers the advantage of a reduced and constant feature space. PC-mer effectively decreases the required number of features for classification, reducing it by around ${2^{k}}$ times in comparison to traditional k-mer-based methods. This reduction in features results in enhanced computational efficiency and efficient utilization of memory, enabling the classifier to handle large-scale virus datasets effectively. In addition, PC-mer, as a k-mer-based method, maintains a constant number of features irrespective of the length of the sequence. This unique property sets it apart from integer and one-hot encoding methods. The constant feature space ensures that the classifier can process sequences of varying lengths without introducing any additional computational overhead.

		\item[2.] Improved Speed and Scalability: PC-mer offers faster processing times and superior scalability compared to traditional alignment methods. It enables researchers to analyze extensive datasets and perform comprehensive comparative studies efficiently. In addition, to use PC-mer in ML-based classifier methods, the reduction in feature space and efficient sequence embedding generation contribute to an average 88\% improvement in training speed compared to traditional k-mer-based methods. This acceleration is especially advantageous when dealing with extensive virus datasets.
		
		\item[3.] Versatility: The PC-mer function is versatile, allowing researchers to explore various distance metrics and adjust the "k" parameter to adapt the analysis to different biological contexts.
		
		\item[4.] Robustness: PC-mer is robust in handling diverse biological datasets, such as those containing rapidly evolving sequences or structural variations, where conventional alignment methods may struggle.
	\end{itemize}
	
	In the subsequent sections, we elaborate on the PC-mer function's application in generating sequence embeddings and its utilization for constructing phylogenetic trees. We then present a comparative study of the alignment-free PC-mer approach against a reference distance matrix derived from sequence alignment, showcasing its effectiveness in accurate phylogenetic analysis for various biological datasets. The PC-mer method has been investigated and studied for the classification of various viruses, including SARS-CoV-2 \cite{pc2}. However, not all of its dimensions, capabilities, and limitations have been fully determined. Therefore, in this study, our aim is to assess its performance on different categories of datasets and analyze its ability to accurately classify and compare DNA/RNA sequences. The remainder of this paper is organized as follows. In Section 2, we provide a brief overview of the PC-mer method, the datasets used in our experiments, along with their characteristics and sources. In this section also we present the methodology employed for evaluating the performance of the PC-mer method on the new datasets. The experimental results and analysis are presented in Section 3, followed by a discussion of the findings in Section 4. Finally, in Section 5, we conclude the paper with a summary of our observations, limitations, and future directions for expanding the knowledge and capabilities of the PC-mer method in sequence comparison and classification.


	\section{Methodology}
	
In this article, we will evaluate the PC-mer method \cite{pcmer, pc2} as a k-mer-based approach. The PC-mer method is an embedding technique that takes a DNA/RNA sequence as input and generates an output vector with a specific size of ${3x2^{k}}$. This vector can be used as input for classification using learning-based methods or as input for evaluating sequence differences/similarities using appropriate metrics. Recently, there has been increased attention on the impact of PC-mer in classification with the aid of artificial intelligence tools. Consequently, our focus in this work will be examining PC-mer as a sequence comparison tool. In this section, we will begin by reviewing the PC-mer embedding method, followed by discussing different metrics for distance calculation. Finally, we will address the datasets used and the evaluation process.
	
	\subsection{PC-mer Sequence Embedding}
	As mentioned in the previous section, the encoding of sequential data plays a vital role in various fields, such as genomics and bioinformatics, where extracting meaningful information from DNA or RNA sequences is crucial. As mentioned earlier, methods based on K-mer counting have become popular in recent years. While these methods extract valuable information from sequences, they also come with certain challenges, as discussed earlier. Addressing these challenges, the PC-mer method has been proposed, which we evaluate and examine in this study. This method is a modification of the FCGR method for counting k-mers, aiming to reduce the dimensions of the resulting vectors by altering the letters. In the following section, we provide a brief overview of the structure of this method \cite{pcmer, pc2}.
The PC-mer function is implemented and follows these steps:
	\begin{itemize}
	\item [1.] Categorize the nucleotides into three sets based on distinct chemical and physical features:
Set 1: Purine nucleotides A and G are represented by the symbol R, and pyrimidine nucleotides C and T (U) are represented by the symbol Y, based on structural similarity.
Set 2: Amino set (A and C) represented by the symbol M, keto set (G and T or U) represented by the symbol K, based on chemical properties.
Set 3: Strong hydrogen bond nucleotides (C and G) represented by the symbol S, weak hydrogen bond nucleotides (A and T or U) represented by the symbol W.
	\item [2.] Create three 1D vectors with a length of ${2^{k}}$, with each set of nucleotide symbols represented by the corresponding corners of the vectors.
	\item[3.] Initialize the Chaos Game Representation (CGR) for each nucleotide as a pointer positioned at the center of the CGR space.
	\item [4.] For each nucleotide in the input sequence, perform the following steps:
i. Calculate the position of the nucleotide symbol on the CGR space using a specific formula, determining the position as the middle point between the current pointer location and the assigned polygon corner within the CGR space.
ii. Update the pointer position based on the calculated CGR position.
iii. Increment the k-mer frequency count for
	\end{itemize}

	We apply the PC-mer function to each sequence in the dataset, using different values of \(k\) to generate sequence embeddings. For each \(k\) value, we obtain a unique embedding that encodes the physicochemical properties of \(k\)-mers in the sequence.
	
	\subsection{Distance Metric Selection}
	Distance Metric Selection

In order to assess the similarity between embeddings, we consider a variety of distance metrics. These metrics serve the purpose of quantifying the dissimilarity or similarity between two embeddings and determining their distance in the feature space. The selected distance metrics, along with their brief descriptions, are as follows \cite{dismet1, dismet2, afprj}:

\begin{itemize}

\item [1.] Cosine Distance (Cosine): The cosine distance measures the angle between two vectors in the feature space, providing a measure of their similarity regardless of their magnitudes.

\item [2.] Hamming Distance (Hamming): The Hamming distance is typically used for comparing binary or categorical data. It counts the number of positions at which the corresponding elements of two vectors differ.

\item [3.] Minkowski Distance (with different orders): The Minkowski distance is a generalized distance metric that encompasses other commonly used metrics such as Euclidean distance (order 2), Manhattan distance (order 1), and Chebyshev distance (order infinity). The order parameter determines the formula used for distance calculation.

\item [4.] Canberra Distance (Canberra): The Canberra distance is suitable for comparing data with varying magnitudes. It considers both the absolute difference and the sum of the magnitudes of the corresponding elements in two vectors.

\item [5.] Bray-Curtis Distance (Bray-Curtis): The Bray-Curtis distance is often used to measure dissimilarity between compositional data, such as abundances. It calculates the dissimilarity by considering the sum of absolute differences and the sum of absolute similarities between corresponding elements of two vectors.

\item [6.] Jaccard Distance (Jaccard): The Jaccard distance is commonly employed for comparing sets or binary data. It determines dissimilarity by dividing the size of the symmetric difference between two sets by the size of their union.

\item [7.] Jensen-Shannon Distance (Jensen-Shannon): The Jensen-Shannon distance is based on the Kullback-Leibler divergence and is often used to compare probability distributions. It represents the symmetrized version of the Kullback-Leibler divergence between two probability distributions.

\item [8.] Normalized Google Distance (NGD): The Normalized Google Distance is a metric originally developed for comparing search terms in Google search queries. It quantifies the semantic similarity between two concepts based on their co-occurrence in Google search results.

\item [9.] Chebyshev Distance (Chebyshev): The Chebyshev distance calculates the maximum difference between the corresponding elements of two vectors. It is useful for comparing data with varying ranges or magnitudes.

	\end{itemize}
More details of these metrics can be seen in the "Distance metrics" section in the supplementary materials.
By employing these diverse distance metrics, we can capture different aspects of dissimilarity and similarity between embeddings, enabling more comprehensive and robust analysis in our study.

\subsection{Datasets}
	Evaluating the accuracy of the distance score in classifying short sequences with an average length ranging from 200 bp to 32 kbp is the main focus of this study. To accomplish this, we investigate nine sets of sequences with diverse sizes and lengths. Among these sets, five have been widely acknowledged as benchmarks based on previous literature. These benchmark datasets, namely 3, 4, 5, 6, and 7, serve as crucial reference points for our evaluation and comparison purposes \cite{fuzzy, vector}. The datasets are as follows:

	\begin{itemize}

		\item[1.] 18S rDNA sequences from 11 Arbuscular mycorrhizal fungi (AMF) isolates
		\item[2.] 24 Eutherian Mammal sequences
		\item[3.] 59 ebolavirus complete genomes
		\item[4.] 38 segment N (neuraminidase) of Influenza A virus
		\item[5.] 16S rDNA sequences from 40 bacterial isolates
		\item[6.] 48 Hepatitis E virus (HEV) whole genomes
		\item[7.] 30 coronavirus and 4 out-group whole genomes
		\item[8.] 41 mammalian mitochondrial genomes (mtDNA)
		\item[9.] 113 Human rhinovirus (HRV) and 3 HEV-C whole genomes
		
	\end{itemize}
	More details of these datasets are described in the results section.

\subsection{Performance Evaluation}

In order to assess the effectiveness of the PC-mer method in sequence comparison, we conducted a thorough evaluation using various experimental setups. Firstly, we applied the PC-mer method with different values of k, representing different k-mer lengths, to the introduced datasets. Additionally, we utilized different distance metrics to generate a distance matrix for each dataset.

Subsequently, to evaluate the performance of the PC-mer method, we employed the produced distance matrices to calculate phylogenetic trees. These trees were then compared with the trees obtained from three other independent alignment methods: FFP \cite{ffp}, alfpy (Word Count) \cite{afprj}, and kmacs \cite{kmacs}, all of which are also based on k-mer techniques. To establish a reference point for the comparison, we selected the Clustal Omega alignment method \cite{co} and considered the trees resulting from its distance matrix as the baseline.

By comparing the phylogenetic trees derived from the PC-mer method with those obtained using alternative alignment methods and the Clustal Omega \cite{co} reference method, we were able to evaluate the performance of the PC-mer approach. The comparison allowed us to assess the accuracy, robustness, and efficiency of the PC-mer method in sequence comparison tasks. To evaluate the performance, we employ two metrics to assess the accuracy and reliability of the methods \cite{afprj}. These performance metrics include:
	\begin{itemize}
		\item \textbf{normalized Robinson-Foulds metric}: The normalized Robinson-Foulds (nRF) metric is widely used in phylogenetics for quantifying the dissimilarity between two phylogenetic trees. It measures the difference in tree topologies by counting the number of unique or shared bipartitions (splits of taxa into two groups) between the trees.
		\item \textbf{normalized Quartet Distance}: The normalized Quartet Distance (nQD) is a statistical measure used in phylogenetics to assess the dissimilarity between two phylogenetic trees.
		It specifically focuses on quartets, which are four-taxon subtrees within a tree. The nQD quantifies the proportion of quartets that differ between two trees, providing a normalized value to express their topological dissimilarity.
	\end{itemize}
	
	Of course, in addition to comparing the trees using these metrics, we also utilize result visualization to assess the classification performance of various methods. To gain a better understanding of how well the methods cluster the sequences, we employ Principal Component Analysis (PCA) on the embeddings, which enables us to generate 3D embeddings that can be plotted.

PCA is a powerful statistical technique employed for dimensionality reduction and data visualization. Essentially, PCA aims to transform a dataset containing potentially correlated variables into a new set of variables called principal components, which are linearly uncorrelated. These principal components capture the most significant variations in the data, facilitating simplified analysis and visualization \cite{pc2}.

	\section{Comparative Study}
In this section, we present two subsections that analyze and compare the performance of different approaches. The first subsection focuses on comparing phylogenetic trees using the metrics mentioned earlier. This analysis provides insights into the effectiveness and accuracy of the tested methods in capturing the underlying evolutionary relationships. 

The second subsection utilizes the PCA (Principal Component Analysis) method to assess and compare various encoding methods. PCA allows for dimensionality reduction and visualization of the encoded data, aiding in identifying patterns and differences among the different approaches. By employing this technique, we gain a comprehensive understanding of the performance and discriminative capabilities (sequences clustering) of the encoding methods evaluated in this study.

	\subsection{Phylogenetic Tree Analysis}
Here, we use phylogenetic trees constructed by the Neighbor-Joining (NJ) method from the Clustal Omega \cite{co} result as our reference method. We then proceed to compare the performance of four different methods, namely PC-mer \cite{pcmer, pc2}, alfpy (Word Count) \cite{afprj}, kmacs \cite{kmacs}, and FFP \cite{ffp}, against this reference method across nine datasets. To assess and quantify the differences, we employ two comparison metrics: normalized Robinson-Foulds distance (nRF) and normalized Quartet Distance (nQD) \cite{afprj}. Through this analysis, we gain valuable insights into the effectiveness and accuracy of these methods in capturing the underlying evolutionary relationships represented by the phylogenetic trees. It should be noted that the comparison with the phylogenetic tree should be considered independent of the ability of the tools in classification. We will evaluate the issue of categorization in the next section. The rest of this section is dedicated to the analysis of the performance of different models on each separate dataset as shown in \hyperref[tab:results1]{Table~\ref{tab:results1}}.\\ 

In addition to the main analysis, there are two additional points worth noting. Firstly, regarding the PC-mer, alfpy (Word Count), and kmacs methods, we explore various sizes of k for these methods. We conduct tests with different values of k and select the first value that produces comparable results with the other methods, ensuring fair and meaningful comparisons.

Secondly, the choice of distance method and the optimization of hyperparameters play a crucial role in the performance of PC-mer and alfpy (Word Count). To maximize their effectiveness, we extensively test and fine-tune the hyperparameters, as well as evaluate different distance metrics. For PC-mer and alfpy (Word Count), the optimal distance metric is chosen based on rigorous experimentation. Specifically, we leverage the Jensen-Shannon divergence for FFP, as it has been proven to be an effective measure in capturing the dissimilarities between phylogenetic trees. As for kmacs, we utilize the default distance metric provided by the tool, which has been widely used and tested in the field. By meticulous selection and optimization of these distance measures and hyperparameters, we ensure the accuracy and reliability of our comparative analysis.\\
	
	\textbf{18S rDNA sequences from 11 Arbuscular mycorrhizal fungi (AMF) isolates}: AMF (Arbuscular mycorrhizal fungi) are symbiotic fungi that infect vascular plants by penetrating the root cells. They form arbuscules and vesicles and belong to the phylum Glomeromycota. AMF help plants capture essential nutrients from the soil, such as phosphorus, sulfur, nitrogen, and micronutrients. Their symbiotic relationship with plants played a crucial role in plant colonization on land and the evolution of vascular plants. Our AMF dataset here consists of divergent sequences from the same family levels such as Glomeraceae and Gigasporaceae, not downer taxa levels \cite{fuzzy}. This issue has shown itself in the average value of the identity of dataset AMF in \hyperref[tab:results1]{Table~\ref{tab:results1}}, which shows that the sequences of this dataset are very different from each other. However, the comparison results of PC-mer (nRF and nQD) show that PC-mer has a better ability to compare divergent sequences than other methods, such as the alignment method, and the result is closer to Clastal Omega than other methods. PC-mer managed to get the lowest nRF and nQD among all the other methods on this dataset with the optimal $ k=6$ using the Euclidean distance metric.   \\
	
	\textbf{24 Eutherian Mammal sequences}: The Eutherian mammal sequence dataset comprises genetic sequences of transferrin and lactoferrin from a diverse set of 24 vertebrate species. Transferrins and lactoferrins are iron-binding proteins that play vital roles in iron storage, as well as providing protection against bacterial diseases. These sequences, collected from different bodily fluids and spaces such as blood serum, milk, eggs, tears, and interstitial spaces, offer valuable insights into the functions and variations of these proteins across vertebrate species \cite{fuzzy}. The Eutherian mammal dataset showed that PC-mer outperformed kmacs in terms of performance metrics such as nQD and nRF. PC-mer achieved a lower nQD value and the same nRF value as kmacs, with an optimal k value of 12 using the Jaccard distance metric. However, when compared to alfpy and ffp, PC-mer exhibited larger nRF and nQD values. It is worth noting that PC-mer achieved these superior results with a smaller feature vector size. PC-mer utilized only ${3\times2^{12}}$ features, whereas alfpy, kmacs, and ffp employed ${4^{12}}$, ${4^{30}}$, and ${4^{30}}$ features, respectively.
 \\

	\textbf{38 segment N (neuraminidase) of Influenza A virus}: Influenza A viruses are known troublemakers, infecting a wide range of hosts including humans. These viruses possess a segmented, single-stranded, negative-sense RNA genome divided into 8 segments. The hemagglutinin (H) and neuraminidase (N) proteins on their viral surfaces exhibit various serotypes \cite{vector, inf}. Our study focused on analyzing the segment N through phylogenetic analysis. Our dataset for Influenza A comprises 38 samples derived from five separate clusters: H1N1, H5N1, H2N2, H7N3, and H7N9. These clusters encompass four distinct types of the neuraminidase gene segment, namely N1, N2, N3, and N9.
PC-mer outperformed all the other methods on this dataset in terms of the nRF, reaching the lowest value of 0.11 with an optimal $k=9$ using Jaccard distance metric.  \\
	
	\textbf{113 Human rhinovirus (HRV) and 3 HEV-C whole genomes}: To evaluate our proposed methodology, we conducted a phylogenetic analysis of Human rhinoviruses (HRVs), which are commonly associated with the occurrence of the common cold and contribute to over half of cold-like illnesses. HRVs are classified as single-stranded positive sense RNA viruses with a length of approximately 7200 nucleotides. The genome of HRVs consists of a single gene that codes for a polyprotein, subsequently cleaved to produce 11 distinct proteins. Belonging to the Enterovirus genus and Picornaviridae family, HRVs exhibit three distinct phylogenetic clusters known as HRV-A, HRV-B, and HRV-C \cite{vector}. Here in addition to utilized 113 complete genomes of these three clusters to explore their evolutionary relationships, we have three HEV-C genomes as outgroups. With an optimal $k=17$ using the Jaccard distance, PC-mer reached the same nRF value as alfpy (Word Count) and kamcs, and a higher value than FFP. The results show us that PC-mer is at least as reliable as kmacs and alfpy.\\
	
	\textbf{48 Hepatitis E virus (HEV) whole genomes}: In this study, we included a set of 48 complete genomes of hepatitis E virus (HEV) to serve as an additional benchmark. HEV is classified as a non-enveloped, single-stranded RNA virus with a genome size of approximately 7200 nucleotides. Notably, the hepatitis E virus is responsible for inducing acute hepatitis in infected individuals. One unique characteristic that distinguishes it from other well-known hepatitis viruses (such as A, B, C, and D) is that it is the only known hepatitis virus that primarily affects animals and can be transmitted to humans \cite{vector, fuzzy}. For this dataset, PC-mer achieved an nRF value of 0.13 using the Jaccard metric, with the optimal k value set at 12. Remarkably, this result is consistent with the value obtained by the FFP method and ranks second among the methods we compared. It is worth mentioning that PC-mer, despite having smaller feature vector sizes than other methods, yielded comparable performance in this case.	\\
	
	\textbf{59 ebolavirus complete genomes}: In this study, we conduct an analysis of a dataset containing complete genomes of the ebolavirus. The genomes were quite lengthy, consisting of approximately 18,900 nucleotides. The ebolavirus genus includes five distinct species: Bundibugyo virus (BDBV), Reston virus (RESTV), Ebola virus (EBOV), Sudan virus (SUDV), and Tai Forest virus (TAFV). Each species has its own unique characteristics. One notable feature of ebolaviruses is their genetic makeup. They are classified as single-strand negative-sense RNA viruses. This means that their genetic material consists of a single RNA strand with an opposite orientation compared to the typical positive-sense RNA. Within the genome of each ebolavirus, there are several encoded proteins. One particularly important protein is the glycoprotein, which is found on the surface of the ebolavirus. It plays a crucial role in the virus's interactions with host cells \cite{vector, fuzzy}. This dataset is one of the hard benchmarks for alignment-free methods. Of course, PC-mer outperforms FFP in terms of the nRF and the nQD, reaching an nRF of 0.39 and an nQD of 0.0434 with an optimal $k=11$ using the Minkowski distance. Our method is also better than alfpy (Word Count) in terms of the nQD. In general, it can be said that this tool is in the second comparison.
	\\
	
	\textbf{30 coronavirus and 4 out-group whole genomes}: Coronaviruses are enveloped, single-stranded, positive-sense RNA viruses belonging to the family Coronaviridae. They have a pleomorphic nature and can infect a wide range of hosts including avians, bats, humans, and other mammals. These viruses have the ability to cause various respiratory, gastrointestinal, neurological, and systemic diseases, ranging from mild to severe. Their unique feature is the capability to cross the species barrier and infect new hosts. In this study, a dataset consisting of 30 complete coronavirus genomes and 4 non-coronaviruses as outgroups was utilized. The 30 coronavirus genomes were categorized into five groups based on their host types \cite{vector, fuzzy}. The analysis of these genomes using our method provided valuable insights into the classification and evolutionary relationships of coronaviruses, particularly in light of past pandemics like SARS. For this dataset, Our method, PC-mer, outperforms all the other ones in terms of the nRF, reaching an nRF of 0.29 with an optimal $k=6$ using the Bray-Curtis distance. PC-mer also reached an nQD of 0.0994, which is lower than the values given by kmacs and FFP.
	\\
	
	\textbf{16S rDNA sequences from 40 bacterial isolates}: this dataset is selected among datasets of fuzzy paper. This dataset focuses on 40 16S rDNA sequences obtained from pure cultures of endophytic bacteria isolated from mature endosperms of six rice varieties collected from distinct locations in north-east India. Utilizing a combination of primer pairs (27f/1492r and 533f/805r), the full-length 16S rDNA sequences were amplified and subsequently sequenced using the BigDye terminator method. To ensure sequencing accuracy, contigs were assembled based on phred scores and screened for potential chimeras. Alignment against the NCBI reference rRNA database using the BLASTN algorithm enabled the identification of the bacterial isolates. The results demonstrated the presence of diverse endophytic bacterial species associated with the investigated rice varieties, highlighting their potential functional roles in plant systems \cite{fuzzy}. As can be observed from the performance of compared methods, comparison and classification of this dataset is very hard. However, PC-mer displays an outstanding performance on this dataset, reaching an nRF of 0.59 and an nQD of 0.2473 with an optimal $k=6$ using the Minkowski Distance. The results are better than all the other methods by far in terms of both the nRF and the nQD which shows that the PC-mer method has the most similar phylogenetic tree to the CLUSTAL Omega method compared to other methods.
	\\
	
	\textbf{41 mammalian mitochondrial genomes (mtDNA)}: The benchmark dataset utilized in this study comprises 41 complete mitochondrial genomes (mtDNA) from various mammalian species, encompassing approximately 16,500 base pairs. mtDNA, a circular and double-stranded genetic material, exhibits a unique structural feature wherein one strand, known as the heavy strand, is rich in guanine, while the other strand, referred to as the light strand, is enriched in cytosine. In this dataset, we focused specifically on the heavy strands of mtDNA. Notably, these sequences exhibit relatively low conservation and display a rapid mutation rate, contributing to their potential as informative markers for evolutionary studies and population genetics analyses \cite{vector, fuzzy}. This dataset represents a highly intricate collection used for comparison and classification tasks. Despite these challenges, the PC-mer approach, characterized by a relatively fewer number of features compared to alternative methods, has achieved an impressive second rank in terms of phylogenetic tree similarity. With an optimal $k=15$ using the Jaccard distance, PC-mer reached an nQD of 0.1563, which is better than that of alfpy(Word Count) and kmacs.

\begin{table}[H]
		\caption{Comparison of Methods based on nRF and nQD Metrics}
		\label{tab:results1}
\begin{center}
		\begin{tabular}{cccccccc}
			\toprule
			\multirow{2}{*}{Method} &  \multirow{2}{*}{k} & \multirow{2}{*}{Dataset} &  {Iden. (Avg.)} & {Iden. (STD)} & \multicolumn{3}{c}{Metrics} \\
			\cmidrule{6-8}
			&&&&& nRF & nQD & Distance Metric \\
			\midrule
			PC-mer & 6 & AMF & 59.06&20.86 & 0.12 & 0.0727 & Euclidean \\
			alfpy (Word Count) & 6 & AMF &&& 0.25 & 0.1636 & Euclidean \\
			kmacs & 10 & AMF &&& 0.25 & 0.1636 & - \\
			FFP & 5 & AMF &&& 0.38 & 0.1212 & - \\
			\midrule
			PC-mer & 12 & Euth. Mammal & 60.44 & 16.30 & 0.24 & 0.0418 & Jaccard \\
			alfpy (Word Count) & 12 & Euth. Mammal &&& 0.19 & 0.034 & Google \\
			kmacs & 30 & Euth. Mammal&& & 0.24 & 0.0489 & - \\
			FFP & 30 & Euth. Mammal &&& 0.19 & 0.0331 & - \\
			\midrule
			PC-mer & 9 & Influenza & 69.49 &16.10& 0.11 & 0.0236 & Jaccard \\
			alfpy (Word Count) & 12 & Influenza &&& 0.2285 & 0.042 & Euclidean \\
			kmacs & 70 & Influenza &&& 0.2 & 0.0364 & - \\
			FFP & 20 & Influenza &&& 0.2 & 0.018 & - \\
			
			\midrule
			PC-mer & 17 & HRV & 65.10&9.54 & 0.14 & 0.0288 & Jaccard \\
			alfpy (Word Count) & 10 & HRV &&& 0.14 & 0.02753 & Google \\
			kmacs & 30 & HRV & 0.14 &&& 0.0276 & - \\
			FFP & 20 &HRV&&& 0.12 & 0.0236 & - \\
			
			\midrule
			PC-mer & 12 & HEV & 78.67 & 8.08 & 0.13 & 0.0054 & Jaccard \\
			alfpy (Word Count) & 7 & HEV &&& 0.15 & 0.006 & Euclidean \\
			kmacs & 20 & HEV &&& 0.11 & 0.002 & - \\
			FFP & 20 &HEV&&& 0.13 & 0.0048 & - \\
			
			\midrule
			PC-mer & 11 & Ebola &73.16 &21.79& 0.39 & 0.0434 & Minkowski(3) \\
			alfpy (Word Count) & 12 & Ebola &&& 0.375 & 0.049 & Google \\
			kmacs & 10 & Ebola &&& 0.38 & 0.0219 & - \\
			FFP & 30 &Ebola&&& 0.41 & 0.0448 & - \\
			
			\midrule
			PC-mer & 6 & Corona & 45.24 & 32.21 & 0.29 & 0.0994 & Bray-Curtis \\
			alfpy (Word Count) & 12 & Corona &&& 0.322 & 0.059 & Euclidean \\
			kmacs & 70 & Corona &&& 0.35 & 0.1654 & - \\
			FFP & 30 &Corona&&& 0.32 & 0.1154 & - \\
			
			\midrule
			PC-mer & 6 & rDNA Bacterial &79.73& 16.60 & 0.59 & 0.2473 & Minkowski(3) \\
			alfpy (Word Count) & 14 & rDNA Bacterial &&& 0.7837 & 0.2856 & Euclidean \\
			kmacs & 30 & rDNA Bacterial &&& 0.95 & 0.4153 & - \\
			FFP & 8 &rDNA Bacterial&&& 0.95 & 0.3358 & - \\
			
			\midrule
			PC-mer & 15 & mtDNA &69.74 &6.73 & 0.26 & 0.1563 & Jaccard \\
			alfpy (Word Count) & 12& mtDNA &&& 0.23 & 0.1573 & Euclidean \\
			kmacs & 10 & mtDNA &&& 0.26 & 0.1716 & - \\
			FFP & 40 &mtDNA& &&0.24 & 0.1448 & - \\
			\bottomrule
		\end{tabular}
\end{center}
		\tablefootnote{the hyperparameters and the distance metrics are tested and optimally chosen for PC-mer and alfpy (Word Count). The Jensen-Shannon divergence is used for FFP, and the default distance metric that comes with the tool is used for kmacs.}
	\end{table}

	\subsection{Evaluating the embedded method}
	In this section, we explored the classification abilities of four encoding methods, namely PC-mer \cite{pcmer,pc2}, FFP \cite{ffp}, Alfpy \cite{afprj}, and kmacs \cite{kmacs}. To assess their effectiveness, we applied Principal Component Analysis (PCA) to reduce the dimensionality of the encoded datasets. Subsequently, we visualized the results using 3D plots, where each data point represented a sample. The different classes or labels were distinguished through the use of color or markers. These visualizations provided insights into the separation and clustering of samples for each encoding method in a visually intuitive manner. It should be noted that this evaluation is qualitative and for a more accurate evaluation, it is necessary to use quantitative methods.
\\

\begin{figure}[H]
		\centering
		\includegraphics[width=\linewidth]{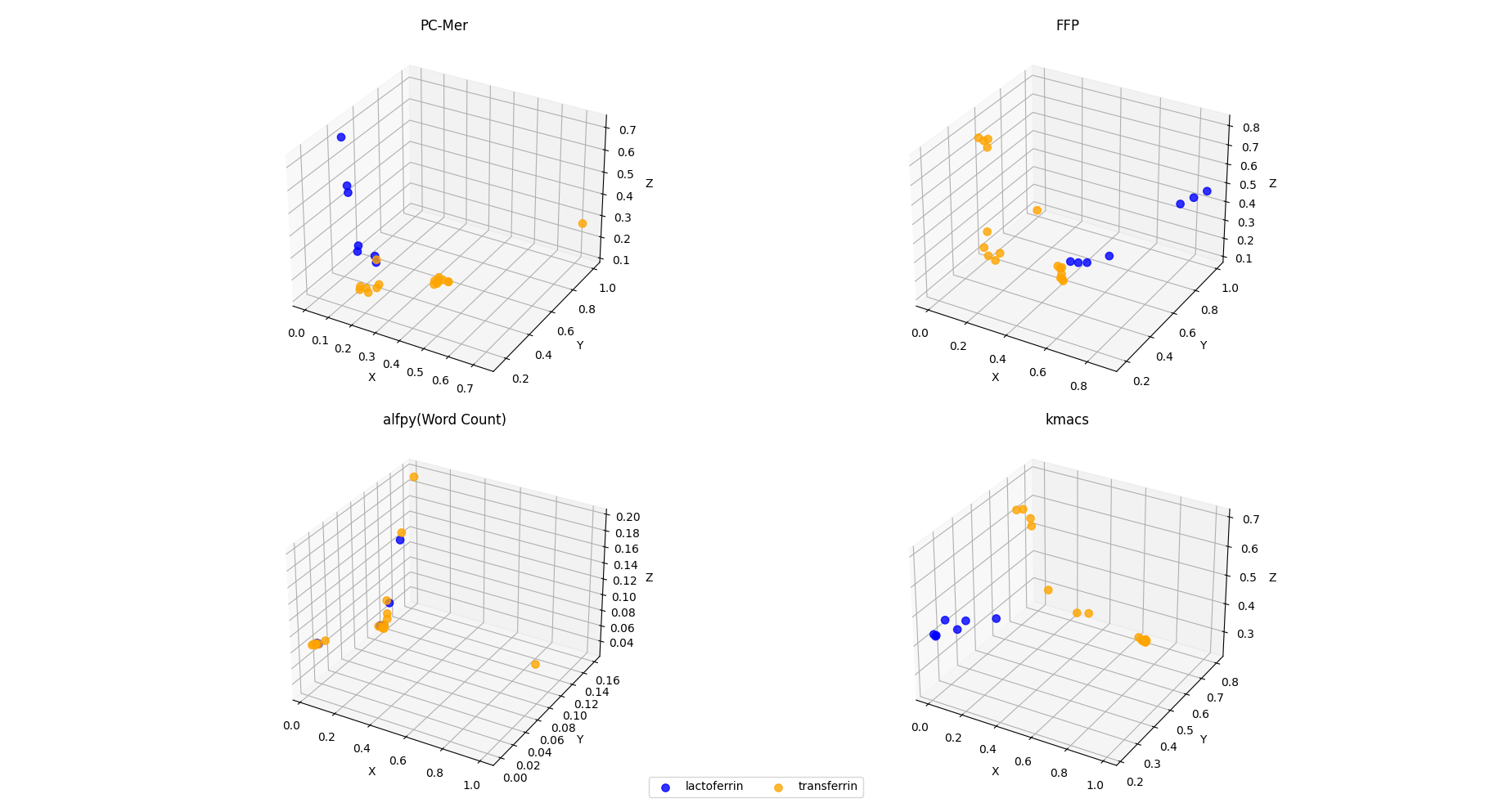}
		\caption{Eutherian Mammal 3D Embeddings}
		\label{fig:em}
	\end{figure}

\textbf{24 Eutherian Mammal sequences}: In the PCA plot (\hyperref[fig: em]{Figure~\ref{fig:em}}), it is evident that the Alfpy encoding method does not effectively separate the two clusters of Eutherian Mammal sequences. The two clusters appear to overlap or mix together, indicating a lack of distinctiveness in the encoded representations. On the other hand, PC-mer and the other encoding methods demonstrate a successful separation of the two clusters. Their encoded representations exhibit clear boundaries, with minimal overlap between the clusters. This suggests that PC-mer, along with the other methods employed, possess superior discriminative capabilities in capturing the underlying patterns and variations within the Eutherian Mammal sequences.
	
\begin{figure}[H]
		\centering
		\includegraphics[width=\linewidth]{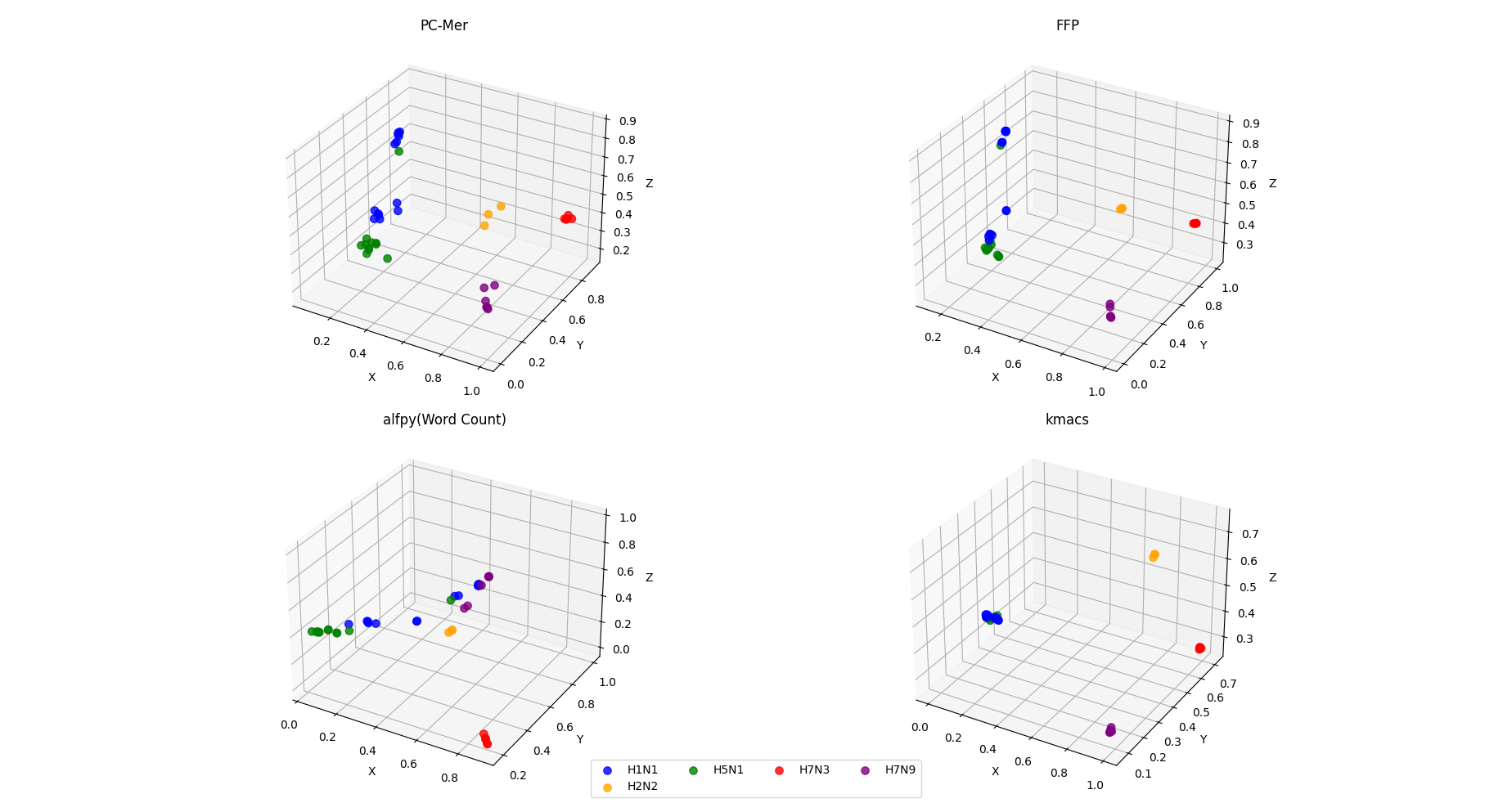}
		\caption{Influenza 3D Embeddings}
		\label{fig:Inf}
	\end{figure}

\textbf{38 segment N (neuraminidase) of Influenza A virus}: Based on the observations from \hyperref[fig: Inf]{Figure~\ref{fig:Inf}}, it can be concluded that PC-mer outperforms the other methods in effectively separating the clusters of the 38 segment N (neuraminidase) of Influenza A virus. The encoded representations generated by PC-mer exhibit distinct boundaries between the clusters, indicating a higher level of discrimination. However, it's worth noting that in terms of clustering, the kmacs method combines the samples from two classes, namely H1N1 and H5N1. This suggests a potential limitation of the kmacs method in accurately clustering the samples based on their distinct characteristics. Therefore, for this specific dataset, PC-mer appears to be the preferred choice as it demonstrates superior performance in both cluster separation and overall clustering accuracy.

\begin{figure}[H]
		\centering
		\includegraphics[width=\linewidth]{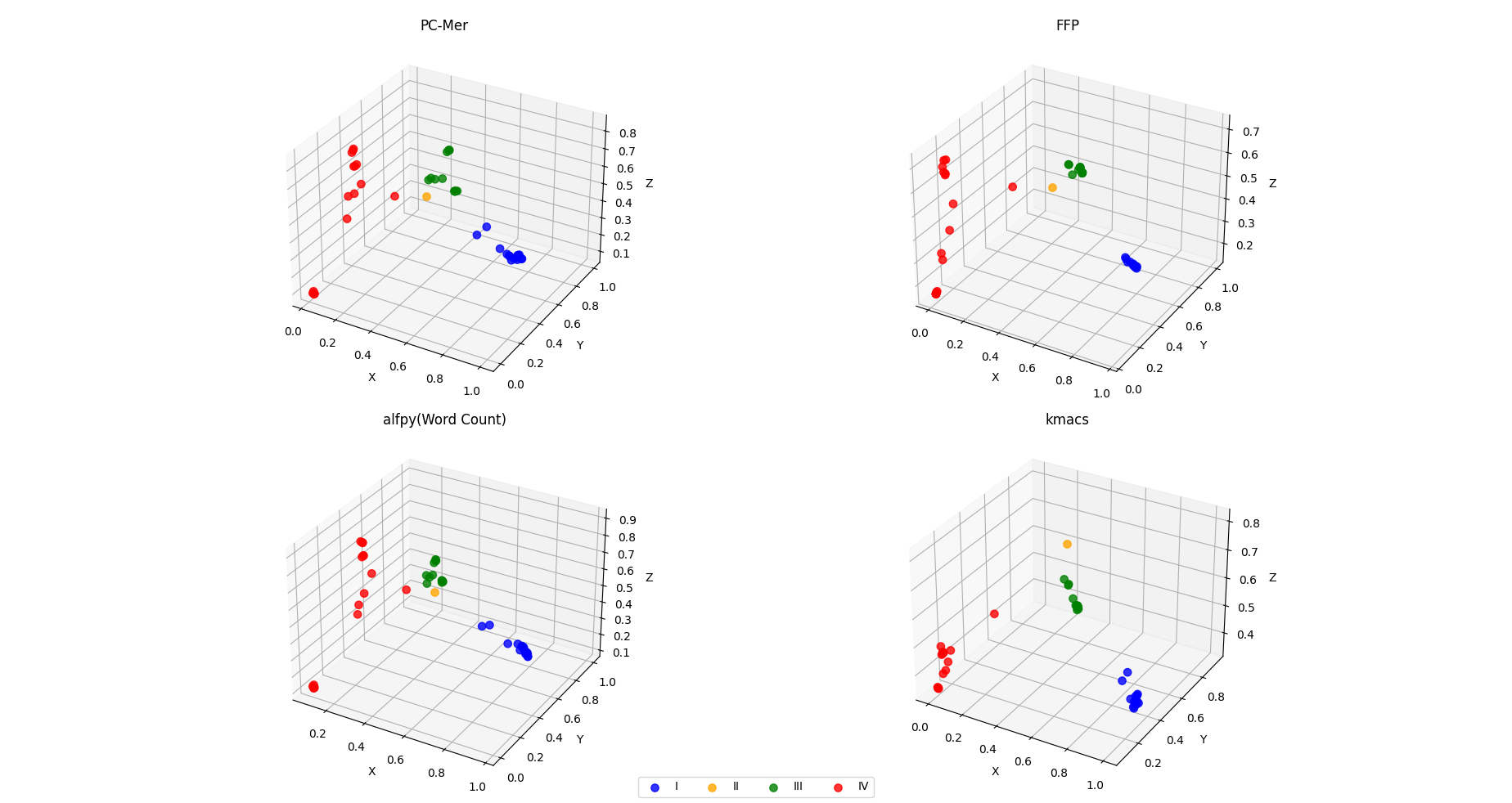}
		\caption{HEV 3D Embeddings}
		\label{fig:hev}
	\end{figure}

\textbf{48 Hepatitis E virus (HEV) whole genomes}: Based on the observations from \hyperref[fig:hev]{Figure~\ref{fig:hev}}, it appears that all the methods used for the dataset of 48 Hepatitis E virus (HEV) whole genomes are capable of separating the samples into four distinct classes. This suggests that each method successfully captures the underlying patterns and features of the HEV genomes, allowing for effective discrimination between different classes. However, without further context or specific quantitative analysis, it is difficult to determine if one method outperforms the others in terms of cluster separation quality or accuracy. Nevertheless, the fact that all methods are able to separate the samples into four distinct classes indicates that they are generally suitable for analyzing and clustering this particular dataset.

\begin{figure}[H]
		\centering
		\includegraphics[width=\linewidth]{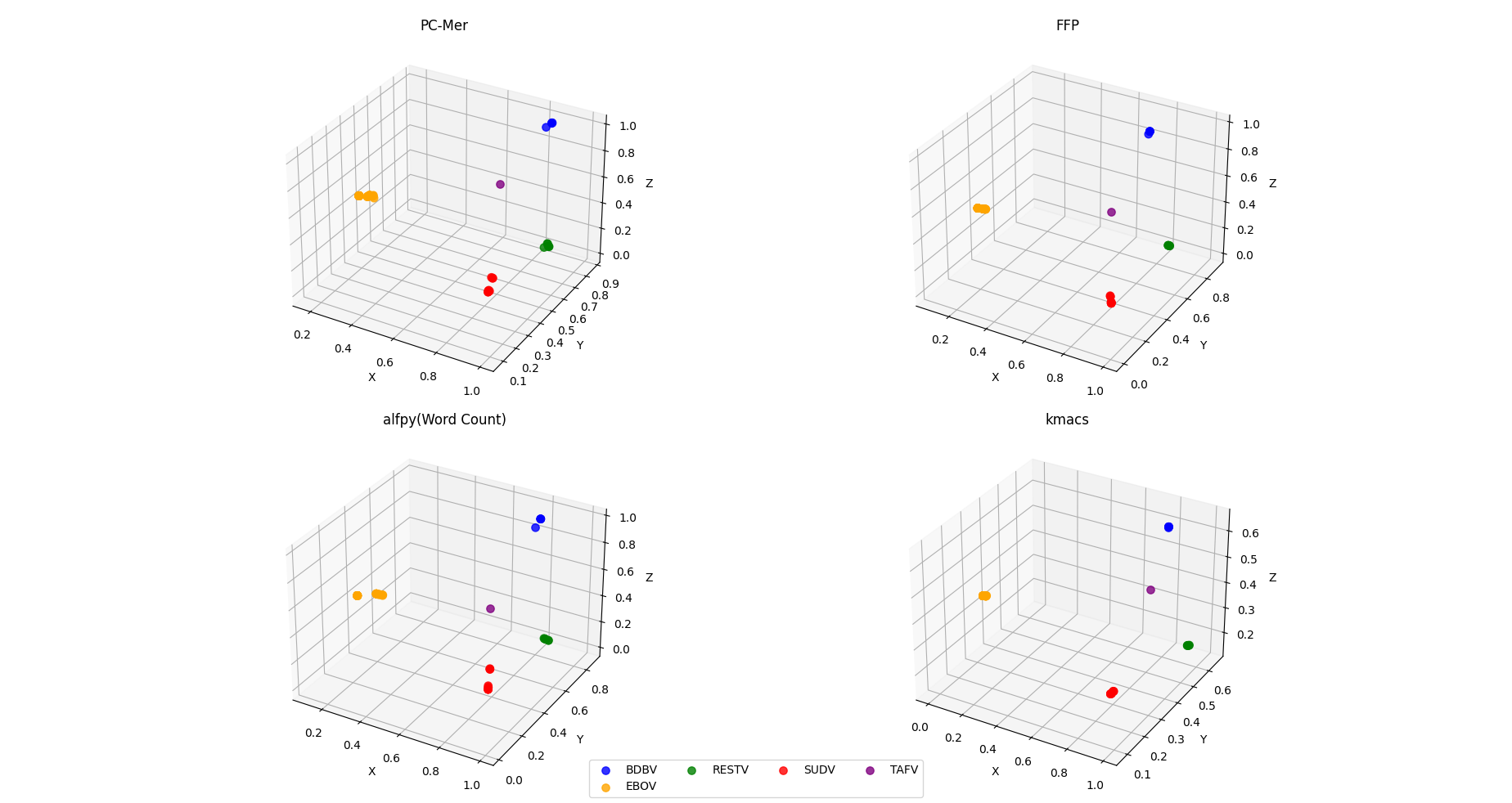}
		\caption{Ebola 3D Embeddings}
		\label{fig:ebola}
	\end{figure}

\textbf{59 ebolavirus complete genomes}: According to \hyperref[fig:ebola]{Figure~\ref{fig:ebola}}, It seems that all the methods used for the dataset of 59 ebolavirus complete genomes perform well. This suggests that each method is effective in capturing the patterns and characteristics within the ebolavirus genomes, leading to accurate analysis and results.
	
\begin{figure}[H]
		\centering
		\includegraphics[width=\linewidth]{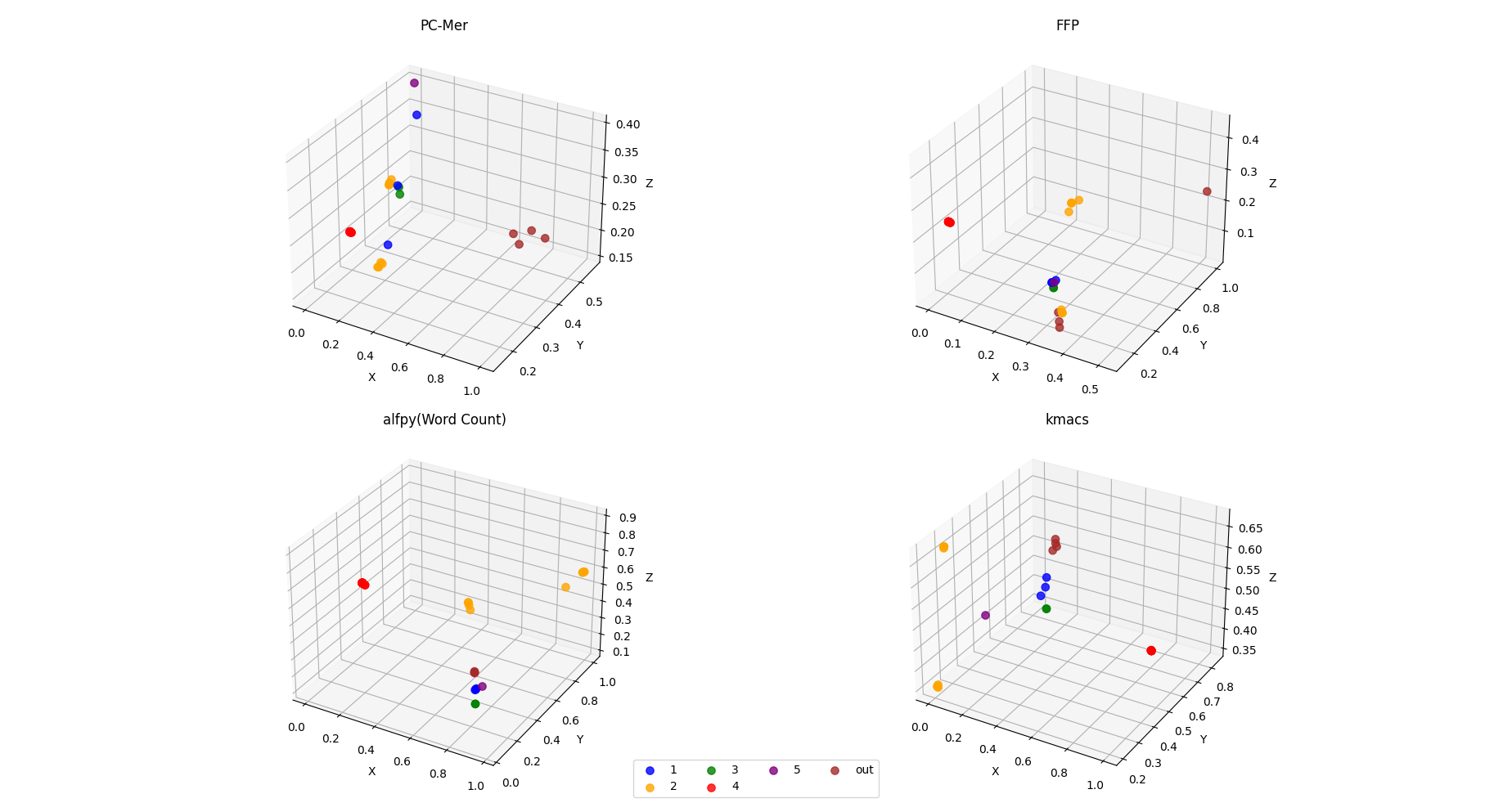}
		\caption{Corona 3D Embeddings}
		\label{fig:corona}
	\end{figure}

\textbf{30 coronavirus and 4 out-group whole genomes}: It appears that clustering the dataset of 30 coronavirus genomes and 4 out-group genomes presents some challenges. However, PC-mer seems to be successful in separating all the clusters except for cluster 1. Additionally, the low intra-cluster distances indicated in the plot suggest that PC-mer is effective in creating distinct clusters within the dataset. Based on these observations, it can be concluded that PC-mer performs better or at least as well as the other methods used in this analysis. This conclusion aligns with the results presented in \hyperref[fig:corona]{Figure~\ref{fig:corona}}, further reinforcing the superiority or comparable capability of PC-mer in clustering the dataset. It is worth noting that without specific information about the other methods and their performance, it's difficult to make a definitive statement about PC-mer being better than all other methods. However, the fact that PC-mer successfully separates most clusters and achieves low intra-cluster distances is a positive indication of its efficacy in this particular dataset. 
	
\begin{figure}[H]
		\centering
		\includegraphics[width=\linewidth]{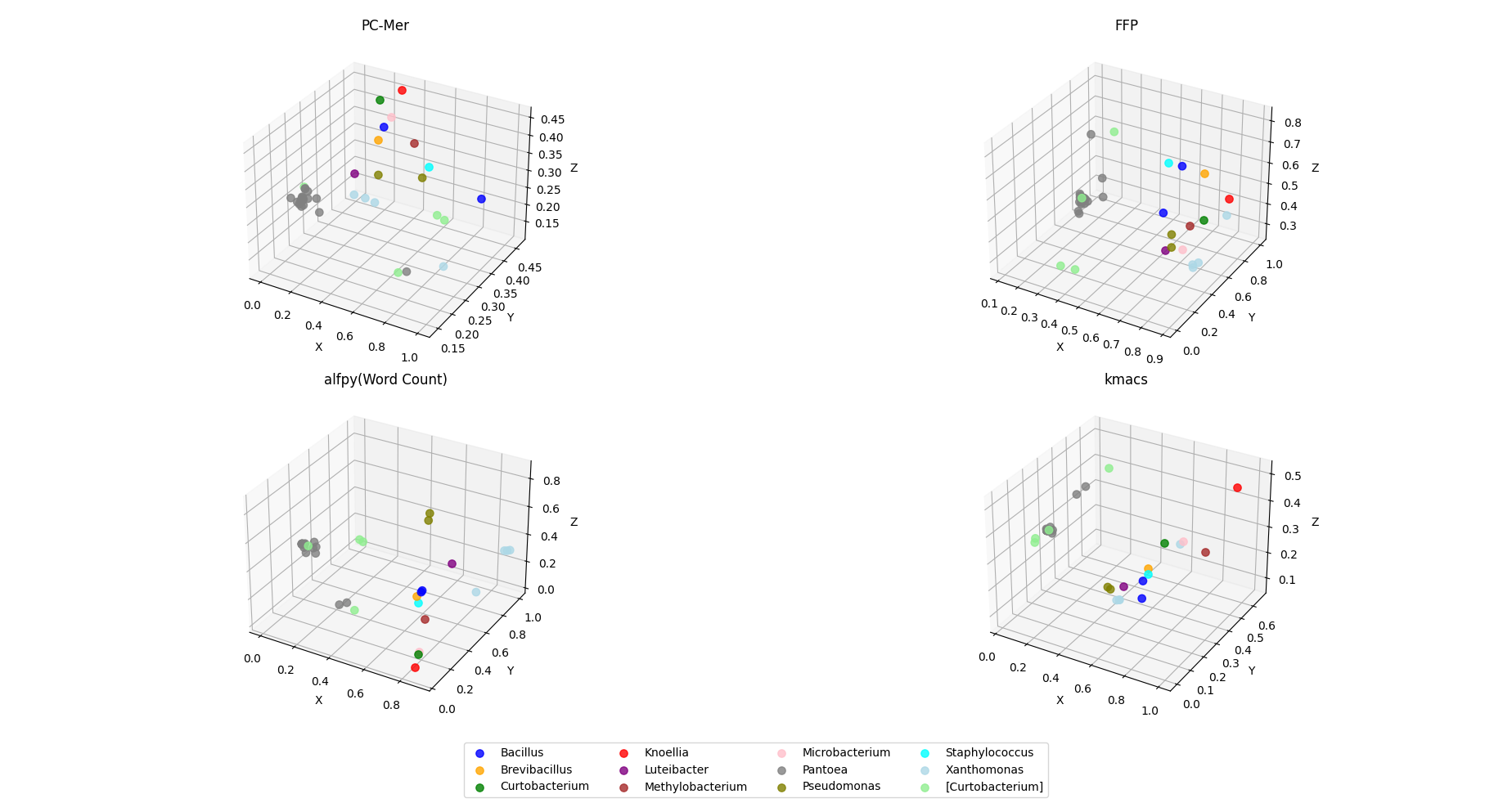}
		\caption{rDNA Bacterial 3D Embeddings}
		\label{fig:rdna}
	\end{figure}

	\textbf{16S rDNA sequences from 40 bacterial isolates}: According to \hyperref[tab:results1]{Table~\ref{tab:results1}}, PC-mer outperforms the other methods in constructing the phylogenetic tree for the dataset of 16S rDNA sequences from 40 bacterial isolates. This indicates that PC-mer is particularly effective in capturing the relationships and clustering patterns within the dataset. Moreover, \hyperref[fig:rdna]{Figure~\ref{fig:rdna}} provides visual evidence supporting the superiority of PC-mer in separating the embeddings belonging to the Pantoea cluster and placing them in close proximity. This is an important observation since it indicates that PC-mer accurately captures the distinctiveness of the Pantoea cluster. 
Additionally, PC-mer is not only successful in the Pantoea cluster but also retains competitive performance for the other clusters within the dataset. Overall, these findings confirm that PC-mer is the preferred method for constructing the phylogenetic tree in this specific dataset, as it outperforms other methods and maintains or exceeds their performance in separating the clusters.

\begin{figure}[H]
		\centering
		\includegraphics[width=\linewidth]{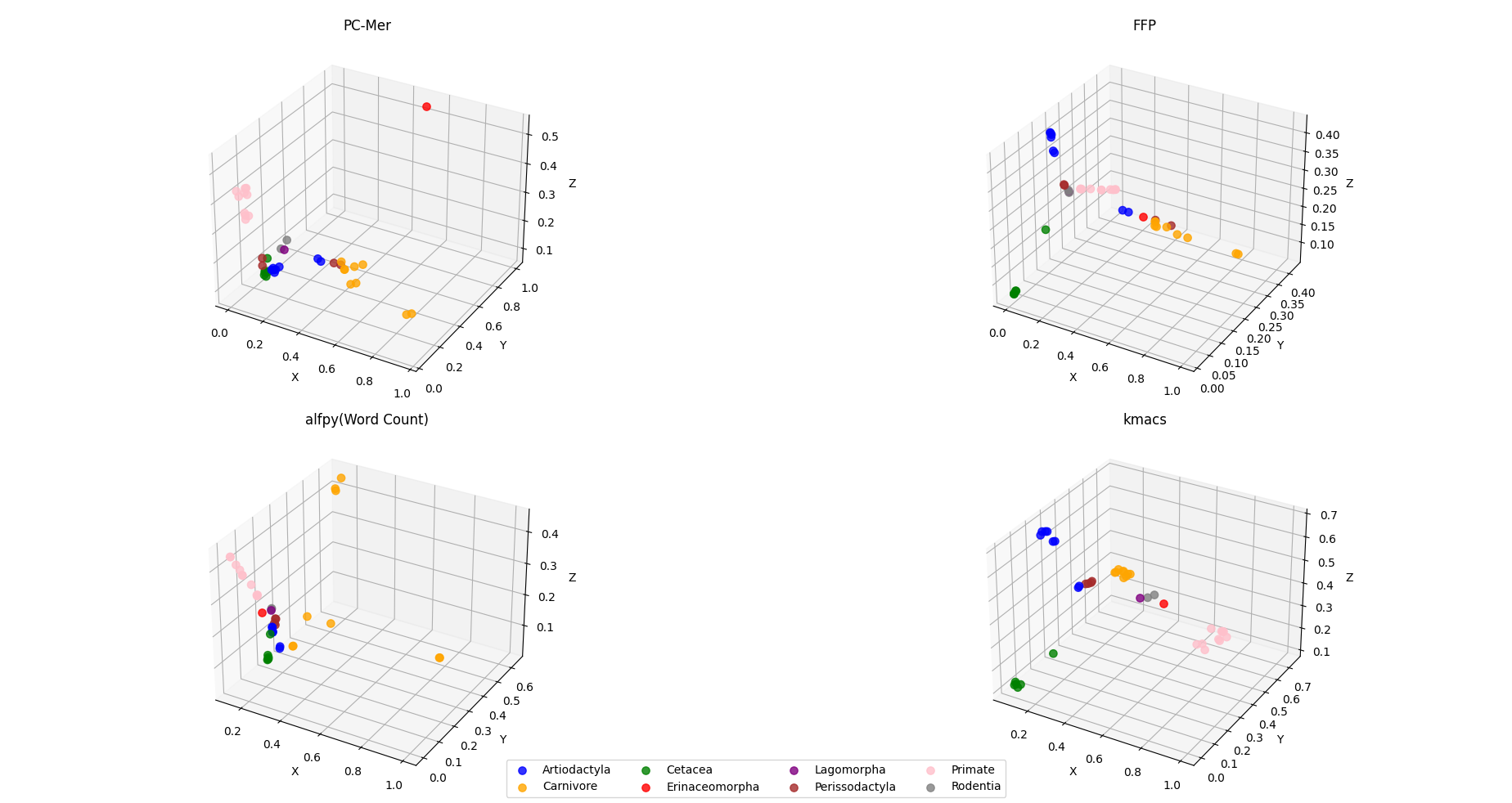}
		\caption{mitochondrial 3D Embeddings}
		\label{fig:mito}
	\end{figure}

	\textbf{41 mammalian mitochondrial genomes (mtDNA)}: \hyperref[fig:mito]{Figure~\ref{fig:mito}} demonstrates that PC-mer outperforms alfpy (Word Count) and FFP in clustering Carnivores and Primates together but separate from other clusters in the dataset of 41 mammalian mitochondrial genomes (mtDNA). This finding highlights the efficacy of PC-mer in capturing the evolutionary relationships between these specific groups.
		
\section{Disscusion and Conslusion}
The results obtained from this study further reinforce the reliability and performance of PC-mer as a clustering method. These findings align with previous research papers that have explored the application of PC-mer in different domains, such as classifying SARS-CoV-2 and viruses in general.

In a previous study \cite{pc2} on classifying SARS-CoV-2 sequences, PC-mer demonstrated a high level of accuracy and robustness. The method effectively captured the genetic variations and evolutionary relationships among different strains of the virus, enabling accurate classification and identification. This success in the classification of SARS-CoV-2 sequences highlights the versatility and effectiveness of PC-mer in handling viral genomic data.

Additionally, a separate paper \cite{pcmer} focused on the classification of diverse viral genomes using PC-mer. The results showcased the superior performance of PC-mer compared to traditional methods, such as sequence alignment methods, in terms of clustering accuracy and speed. PC-mer proved to be a reliable tool in discerning the genetic diversity and evolutionary patterns within viral populations, leading to valuable insights for surveillance and epidemiological studies.

Building upon the achievements of these previous papers, the present study further verifies the efficacy of PC-mer in clustering different datasets. The method consistently outperformed other approaches, exhibiting lower normalized Robinson-Foulds (nRF) and normalized Quartet Distance (nQD) values on the most of datasets. Additionally, the plots displayed the ability of PC-mer to effectively separate clusters within the tested datasets, emphasizing its discriminative power.

In conclusion, the results from this study, along with the outcomes of previous papers on SARS-CoV-2 classification and viral genome classification \cite{pcmer, pc2}, collectively reinforce the robustness and reliability of PC-mer as a clustering method. Its accurate classification and separation capabilities make it a valuable tool in various domains, ranging from viral genomics to broader sequence analysis. Further research and application of PC-mer hold significant promise in advancing our understanding of genetic diversity and evolutionary relationships across different datasets. 

	\newpage
	\bibliographystyle{ieeetr}
	\bibliography{sample}

\begin{thebibliography}{10}

\bibitem{afprj}
A.~Zielezinski, S.~Vinga, J.~Almeida, and W.~M. Karlowski, ``Alignment-free
  sequence comparison: benefits, applications, and tools,'' {\em Genome
  Biology}, vol.~18, p.~186, Oct. 2017.

\bibitem{vector}
Y.~Li, L.~He, R.~Lucy~He, and S.~S.-T. Yau, ``A novel fast vector method for
  genetic sequence comparison,'' {\em Scientific Reports}, vol.~7, p.~12226,
  Sept. 2017.

\bibitem{fuzzy}
A.~K. Saw, G.~Raj, M.~Das, N.~C. Talukdar, B.~C. Tripathy, and S.~Nandi,
  ``Alignment-free method for dna sequence clustering using fuzzy integral
  similarity,'' {\em Sci. Rep.}, vol.~9, p.~3753, Mar. 2019.

\bibitem{walkim}
S.~Akbari Rokn~Abadi, A.~Mohammadi, and S.~Koohi, ``Walkim: Compact image-based
  encoding for high-performance classification of biological sequences using
  simple tuning-free cnns,'' {\em PLOS ONE}, vol.~17, pp.~1--26, 04 2022.

\bibitem{cvtree}
G.~Zuo and B.~Hao, ``Cvtree3 web server for whole-genome-based and
  alignment-free prokaryotic phylogeny and taxonomy,'' {\em Genomics,
  Proteomics \& Bioinformatics}, vol.~13, no.~5, pp.~321--331, 2015.
\newblock SI: Metagenomics of Marine Environments.

\bibitem{cafe}
Y.~Y. Lu, K.~Tang, J.~Ren, J.~A. Fuhrman, M.~S. Waterman, and F.~Sun, ``Cafe:
  accelerated alignment-free sequence analysis,'' {\em Nucleic Acids Research},
  vol.~45, pp.~W554--W559, 05 2017.

\bibitem{met}
A.~Fiannaca, L.~La~Paglia, M.~La~Rosa, G.~Lo~Bosco, G.~Renda, R.~Rizzo,
  S.~Gaglio, and A.~Urso, ``Deep learning models for bacteria taxonomic
  classification of metagenomic data,'' {\em BMC Bioinformatics}, vol.~19,
  p.~198, jul 2018.

\bibitem{vc1}
A.~Fabijańska and S.~Grabowski, ``Viral genome deep classifier,'' {\em IEEE
  Access}, vol.~7, pp.~81297--81307, 2019.

\bibitem{vc2}
A.-C. Pineda-Peña, N.~R. Faria, S.~Imbrechts, P.~Libin, A.~B. Abecasis,
  K.~Deforche, A.~Gómez-López, R.~J. Camacho, T.~de~Oliveira, and A.-M.
  Vandamme, ``Automated subtyping of hiv-1 genetic sequences for clinical and
  surveillance purposes: Performance evaluation of the new rega version 3 and
  seven other tools,'' {\em Infection, Genetics and Evolution}, vol.~19,
  pp.~337--348, 2013.

\bibitem{pcmer}
S.~A.~R. Abadi, A.~Mohammadi, and S.~Koohi, ``An automated ultra-fast,
  memory-efficient, and accurate method for viral genome classification,'' {\em
  Journal of Biomedical Informatics}, vol.~139, p.~104316, 2023.

\bibitem{mldspsars}
G.~S. Randhawa, M.~P.~M. Soltysiak, H.~El~Roz, C.~P.~E. de~Souza, K.~A. Hill,
  and L.~Kari, ``Machine learning using intrinsic genomic signatures for rapid
  classification of novel pathogens: Covid-19 case study,'' {\em PLOS ONE},
  vol.~15, pp.~1--24, 04 2020.

\bibitem{pc2}
S.~Akbari Rokn~Abadi, A.~Mohammadi, and S.~Koohi, ``A new profiling approach
  for dna sequences based on the nucleotides' physicochemical features for
  accurate analysis of sars-cov-2 genomes,'' {\em BMC Genomics}, vol.~24,
  p.~266, May 2023.

\bibitem{dinuc}
S.~Kariin and C.~Burge, ``Dinucleotide relative abundance extremes: a genomic
  signature,'' {\em Trends in Genetics}, vol.~11, pp.~283--290, jul 1995.

\bibitem{tri}
J.~Zhou, P.~Zhong, and T.~Zhang, ``A novel method for alignment-free dna
  sequence similarity analysis based on the characterization of complex
  networks,'' {\em Evolutionary Bioinformatics}, vol.~12, p.~EBO.S40474, 2016.
\newblock PMID: 27746676.

\bibitem{fcgr}
D.~Lichtblau, ``Alignment-free genomic sequence comparison using fcgr and
  signal processing,'' {\em BMC Bioinformatics}, vol.~20, p.~742, dec 2019.

\bibitem{dft}
T.~Farkaš, J.~Sitarčík, B.~Brejová, and M.~Lucká, ``Swspm: A novel
  alignment-free dna comparison method based on signal processing approaches,''
  {\em Evolutionary Bioinformatics}, vol.~15, p.~1176934319849071, 2019.
\newblock PMID: 31210725.

\bibitem{dismet1}
K.~Chomboon, P.~Chujai, P.~Teerarassamee, K.~Kerdprasop, and N.~Kerdprasop,
  ``An empirical study of distance metrics for k-nearest neighbor algorithm,''
  in {\em Proceedings of the 3rd international conference on industrial
  application engineering}, vol.~2, 2015.

\bibitem{dismet2}
K.~Belattar and S.~Mostefai, ``Similarity measures for content-based
  dermoscopic image retrieval: A comparative study,'' in {\em 2015 First
  International Conference on New Technologies of Information and Communication
  (NTIC)}, pp.~1--6, 2015.

\bibitem{ffp}
G.~E. Sims, S.-R. Jun, G.~A. Wu, and S.-H. Kim, ``Alignment-free genome
  comparison with feature frequency profiles (ffp) and optimal resolutions,''
  {\em Proceedings of the National Academy of Sciences}, vol.~106, no.~8,
  pp.~2677--2682, 2009.

\bibitem{kmacs}
C.-A. Leimeister and B.~Morgenstern, ``{ kmacs: the k -mismatch average common
  substring approach to alignment-free sequence comparison },'' {\em
  Bioinformatics}, vol.~30, pp.~2000--2008, 05 2014.

\bibitem{co}
F.~Sievers, A.~Wilm, D.~Dineen, T.~J. Gibson, K.~Karplus, W.~Li, R.~Lopez,
  H.~McWilliam, M.~Remmert, J.~S{\"o}ding, J.~D. Thompson, and D.~G. Higgins,
  ``Fast, scalable generation of high-quality protein multiple sequence
  alignments using clustal omega,'' {\em Mol. Syst. Biol.}, vol.~7, p.~539,
  Oct. 2011.

\bibitem{inf}
D.~Lichtblau, ``Alignment-free genomic sequence comparison using {FCGR} and
  signal processing,'' {\em BMC Bioinformatics}, vol.~20, p.~742, Dec. 2019.

\end{thebibliography}

\end{document}